\documentclass[12pt,a4paper]{article}
\usepackage[english]{babel}
\usepackage{graphicx,latexsym}

\newcommand{\alt}{\mathbin{\lower 3pt\hbox
   {$\rlap{\raise 5pt\hbox{$\char'074$}}\mathchar"7218$}}}
\newcommand{\agt}{\mathbin{\lower 3pt\hbox
   {$\rlap{\raise 5pt\hbox{$\char'076$}}\mathchar"7218$}}}

\begin{document}
\setcounter{footnote}{0}
\setcounter{equation}{0}
\setcounter{figure}{0}
\setcounter{table}{0}
\vspace*{5mm}



\begin{center}
{\large\bf Triviality, Renormalizability and Confinement
}

\vspace{4mm}
I. M. Suslov \\
Kapitza Institute for Physical Problems,
\\ 119337 Moscow, Russia  \\
 E-mail: suslov@kapitza.ras.ru
\\
\vspace{1mm}
\end{center}

\begin{center}
\begin{minipage}{135mm}
{\bf Abstract } \\
According to recent results, the Gell-Mann -- Low function
$\beta(g)$ of four-dimensional $\phi^4$ theory is non-alternating
and has a linear asymptotics at infinity. According to the
Bogoliubov and Shirkov classification, it means possibility
to construct the continuous theory with finite interaction at
large distances. This conclusion is in visible contradiction
with  the lattice results indicating  triviality of $\phi^4$
theory.  This contradiction is resolved by a special character of
renormalizability in $\phi^4$ theory:  to
obtain the continuous renormalized theory,
there is no need
to eliminate a lattice from the bare theory.
In fact, such kind of renormalizability
is not accidental
and can be understood in the
framework of Wilson's many-parameter renormalization group.
Application of these ideas to QCD shows  that  Wilson's theory
of confinement is not purely illustrative,
but has a direct relation to a real situation.  As a result, the
problem of  analytical proof of  confinement and a mass gap
can be considered as solved, at least on the physical level of
rigor.

\end{minipage}
\end{center}

\vspace*{1.5mm}
\vspace*{1.5mm}

\vspace{3mm}
\begin{center}
{\bf  Introduction }
\end{center}

Recent investigations of the strong coupling regime in $\phi^4$
theory revealed unexpected feature in its renormalizability: the
continual limit in the renormalized theory does not require the
continual limit in the bare theory. We show below that  such kind
of renormalizability has a general character and can be understood
in the framework of Wilson's many-parameter renormalization group.
These results allow to give a final solution to the problem of
triviality or non-triviality of $\phi^4$ theory. Application of
these ideas to the Wilson theory of confinement shows that this
theory is not purely illustrative, but has a direct relation to
real QCD.  As a result, the problem of analytical proof of
confinement and a mass gap can be considered as solved, at least
on the physical level of rigor.

\vspace{3mm}
\begin{center}
{\bf 1. Triviality }
\end{center}

According to recent results \cite{1,2,3,4} (see also \cite{5,6}),
 the Gell-Mann -- Low
function $\beta(g)$ in four-dimensional $\phi^4$ theory is
non-alternating and has asymptotic behavior  $\beta(g)=4 g$
at $g\to\infty$. According to the Bogoliubov and Shirkov
classification \cite{7} (see discussion in \cite{3}), it means
possibility to construct the continuous theory with finite
interaction at large distances.  This conclusion is in visible
contradiction with lattice results indicating triviality of
$\phi^4$ theory (see \cite{8}--\cite{12} and numerous references
in \cite{12a}).

In fact, one should differ two definitions of triviality.
According to Wilson \cite{8}, triviality means that integration
of the Gell-Mann -- Low equation
$$
-\frac{d g}{d \ln L} =\beta(g)
\eqno(1)
$$
 in the direction of large
distances $L$ gives the effective charge $g$ tending to zero
(Fig.1,a); this definition implies the massless theory, since in
the opposite case the distance scale is saturated by the inverse
mass.  The definition of true triviality is different (Fig.1,b).
\begin{figure}
\centerline{\includegraphics[width=5.1 in]{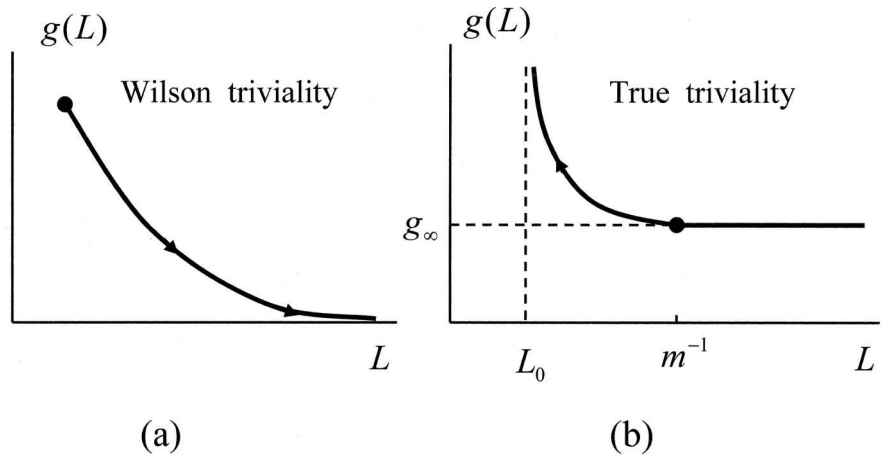}}
\caption{Wilson triviality  (a) against  true triviality
(b). } \label{fig1}
\end{figure}
In this case one considers the massive theory and suggests the
finite interaction $g_{\infty}$ for $L\agt m^{-1}$; a theory is
trivial, if integration of the Gell-Mann -- Low equation in the
direction of small $L$ gives a divergency at finite $L_0$ (the so
called Landau pole) and does not allow to reach the $L\to 0$
limit.  Such situation is internally inconsistent \cite{7} and
means incorrectness of the initial suggestion on finite
interaction at large distances; in fact, $L_0\to 0$ if
$g_{\infty}\to 0$. Wilson triviality means that $\beta$-function
is non-negative and has a zero only for $g=0$. True triviality
needs in addition its sufficiently quick growth at infinity,
$\beta(g)\sim g^\alpha$ with $\alpha>1$. According to
\cite{1}--\cite{6},  $\phi^4$ theory and QED are trivial in the
Wilson sense, but do not possess true triviality.

\vspace{2mm}

Two definitions of triviality were hopelessly
mixed in  literature \cite{12a}. The reasons for it are as
follows:

(a) Bogoliubov and Shirkov's work is poorly known to Western
community;

(b) It is rather difficult to test true triviality in
the lattice approach\,\footnote{\,A definition of
true triviality in the lattice approach was
given in mathematical papers  \cite{9,10}. When the lattice
spacing  $a$ tends to zero, the bare parameters $g_0$
and $m_0$ should be considered as functions of $a$.  A theory is
non-trivial, if there exists some choice of functions $g_0(a)$ and
$m_0(a)$, providing finite interaction at large distances; if
such functions do not exist, then a theory is trivial. Of course,
it is rather difficult to test "existence" or "non-existence" in
numerical simulations.};

(c) There exist arguments that "prove" equivalence of
two definitions.
\vspace{2mm}

\noindent
As illustration  to the latter point, consider the following
reasoning.  The only alternative to perturbative approach is to
express all quantities related to renormalized theory in terms
of the functional integrals. The latter depend on the
bare charge $g_0$, bare mass $m_0$ and the ultraviolet cut-off
$\Lambda$. Taking into account their dimensional
character, one has the following relations for the
renormalized charge $g$, renormalized mass $m$ and
observable quantities $A_i$
$$
g=F_g\left( g_0, m_0/\Lambda\right)\,,\qquad
$$
$$
m=\Lambda F_m\left( g_0, m_0/\Lambda\right)\,,\qquad
\eqno(3)
$$
$$
A_i=\Lambda^{d_i} F_i\left( g_0, m_0/\Lambda\right)\,,\qquad
$$
where $d_i$ is a physical dimensionality of $A_i$.
Excluding  $g_0$ and $m_0/\Lambda$ in favor of $g$ and
$m/\Lambda$, one has
$$
A_i=m^{d_i} \tilde F_i\left( g, m/\Lambda\right)\,.\qquad
\eqno(4)
$$
To eliminate the dependence on $\Lambda$ we should take the
limit $m/\Lambda\to 0$. In the lattice approach, this limit
corresponds to $\xi/a\to\infty$  ($\xi$ is a correlation
length and $a$ is a lattice spacing), i.e. to the phase
transition point. The latter is determined by a zero of
$\beta$-function, which gives $g=0$  in four-dimensional
$\phi^4$ theory.

In this argumentation,  Wilson triviality was considered as
given, while  true triviality was "derived" from it. Of course,
it cannot be correct, because two definitions are surely not
equivalent.  This shortcoming originates from our assumption that
a  general-position situation takes place in Eq.\,4:  in this
case we indeed should take a limit $m/\Lambda\to 0$.  This limit
is unnecessary, if dependence on $m/\Lambda$ is absent in Eq.\,4.
Such special case fills the "gap" between two
definitions and makes them not equivalent.

Such special case really holds in $\phi^4$ theory
\cite{3,4}. Let us return to Eqs.\,3 and impose
the condition $m \ll \Lambda$, corresponding to the continuum
limit of the renormalized theory. If this condition is imposed
in the region $g_0\gg 1$, then  $\phi^4$
theory reduces to the
Ising model, containing the single parameter  $\kappa$,
which plays the role of inverse temperature \cite{3,4};
relations  (3) accept the form
$$
g=F_g\left( \kappa\right)\,,\qquad
$$
$$
m=\Lambda F_m\left( \kappa\right) \,,\qquad
\eqno(5)
$$
$$
A_i=\Lambda^{d_i} F_i\left(\kappa\right)\,. \qquad
$$
So far there is nothing unusual: the condition $m/\Lambda\to 0$
gives a relation between $g_0$ and $m_0/\Lambda$, so all functions
in Eq.3 depend on the single  parameter, which we denoted as
$\kappa$. The non-trivial point consists in the following: the
condition $m/\Lambda\ll 1$ is sufficient for transformation to the
Ising model, but not necessary for it.  In fact, such
transformation is possible under the weaker conditions, which are
compatible with the arbitrary value of $m/\Lambda$ \cite{3,4}.
Excluding $\kappa$ from (5), one obtains the equations
$$
A_i=m^{d_i}  F_i\left( g\right)\,,\qquad
\eqno(6)
$$
which are analogous to (4), but {\it do not contain} the
parameter $m/\Lambda$. As a result, the program of
renormalization is completely fulfilled, and no additional
limiting transitions are necessary. It means that
(a) we can retain the lattice in the bare theory (as
a convenient tool for representation of functional
integrals), and (b) relation between  $m$ and $\Lambda$
(or $\xi$ and $a$) can be arbitrary, so the arbitrary
value of $g$ becomes possible.

Usually, the lattice theory  contains more parameters than
the initial field theory. For example, discretization of the
gradient term in $d$-dimensional $\phi^4$ theory
$$
\int d^dx \left[\nabla\phi(x) \right]^2=
-\int d^dx \phi(x)\nabla^2\phi(x)
 \quad\longrightarrow \quad  \sum_{ x, x'} J_{x-x'}
\phi_{ x}\phi_{ x'} \,,
\eqno(7)
$$
corresponds to the
replacement of $-\nabla^2=\hat p^2$ by
$\epsilon({\hat p})$, where $\epsilon({p})$ is a bare
lattice spectrum
$$
\epsilon({ p})= \sum_{ x} J_{ x} {\rm e}^{i{ p\cdot x}}
= p^2 + O(p^4)   \,,
\eqno(8)
$$
while ${ \hat p}$ is the momentum operator and
$\exp\{i{ \hat p\cdot x}\}$ is the operator of shift on the
vector  ${ x}$. The overlap integrals $J_{ x}$ can be taken
arbitrary and are restricted only by the condition (8).
The interesting question arises: if we can retain a lattice
in the bare theory, then what lattice model should be chosen?

A solution can be found from Eq.\,4. Since dependence on
$m/\Lambda$ is absent, we can take $m/\Lambda\to 0$. But in this
limit (when $\xi/a\to \infty$) there are physical grounds for
independence of functions $F_i$ on the way of cut-off. If such
independence takes place for $m/\Lambda\to 0$, it retains
for arbitrary $m/\Lambda$ due to independence of functions $F_i$
on this parameter. In fact, this argumentation implies
renormalizability of theory (due to which the dependence on
$\Lambda$ can be excluded) and belonging of the lattice model to
the proper universality class (inside of which the dependence on
the way of cut-off is absent).

The lattice theory is frequently considered as a reasonable
approximation to the true field theory.  In this case we should
accept the condition $ \xi \gg a$, which signifies that one has a
lot of lattice sites on the characteristic scale of variation of
field. This condition can be strengthen till $\xi/a\to \infty$ or
liberalized till $ \xi \agt a$. The first case corresponds to  the
point of phase transition and  gives $g=0$. In the second case we
obtain restriction $g\alt 1$ (for the proper charge normalization
 \cite{4}), which can be used to obtain the upper bound on
the Higgs mass \cite{12,13}.

In fact,  the lattice theory should not be considered as any
approximation to field theory, though it is possible for $g_0\ll
1$. The true field theory is continuous from the very beginning
and does not contain any lattice. The lattice is present only in
the bare theory, which is an auxiliary construction and is
completely removed later. No physical requirements, like $\xi\gg
a$,  are relevant for it.  If one removes the condition $\xi\gg
a$, then any values of $g$ become admissible\,\footnote{\,One can
consider $\xi\gg a$ as a technical condition providing a good
approximation, but it is not actual due to the absence of $\xi/a$
dependence.  The stated point of view is in complete agreement
with mathematical definitions  \cite{9,10}, according to which the
limit $a\to 0$ is taken for the arbitrarily chosen dependence
$g_0(a)$ and $m_0(a)$ (see Footnote 1). We impose conditions
$g_0\to\infty$, $g_0^{-1/2} m_0^2 a^2 \to -\infty$, $g_0^{-1}
m_0^2 a^2 =-\kappa$, necessary for transformation to the Ising
model \cite{3,4}.  }. In fact, a real designation of the bare
theory is to represent the relations between physical quantities
in the parametric form (3). Such representation has no deep sense
already due to its ambiguity: it can be written in many different
forms, changing $g_0$ and $m_0/\Lambda$ by any other pair of
variables.

\vspace{2mm}

We see that contradiction between the continual and lattice
approaches is resolved by a special character of
renormalizability in $\phi^4$ theory:
\vspace{2mm}

\noindent
{\it Correct relations (6) between physical quantities
can be obtained for the arbitrary value of the  parameter
$a/\xi$, while the dependence on this parameter is absent.
To obtain the continuous renormalized theory, there is
no need to eliminate a lattice from the bare theory. }

\vspace{3mm}
\begin{center}
{\bf 2. Renormalizability }
\end{center}

The interesting question arises: Is such kind of
renormalizability related with the specific properties of
$\phi^4$ theory,  or it is a manifestation of some general
mechanism?

We shall see below that the second variant is correct. It can be
understood in the framework of Wilson's many-parameter
renormalization group (RG)  \cite{8}.
According to it, the parameters $p_i$
of some lattice Hamiltonian are considered as
functions of the length scale
${\it l}$.\,\footnote{\,Physically it is explained by the well-known
Kadanoff construction. In the description of magnetics,
one begins with the microscopic Hamiltonian for elementary
spins in the lattice sites. Then it is possible to introduce
the macroscopic spin variables corresponding to the blocks of size
${\it l}$ and  write the effective exchange Hamiltonian for
them.  Since the blocks of size  $n{\it l}$ can be composed of
$n^d$ blocks of size ${\it l}$, then recalculation
$p_i({\it l}) \to p_i(n{\it l})$ is possible, i.e.
$p_i(n{\it l})=H_i\left(n,\left\{p_k({\it l})\right\}\right) $.
Taking $n$ close to unity, one can obtain Eqs.9.}
The flow of
these parameters is determined by the RG equations, which can be
written in the differential form
$$
-\frac{d p_i}{d \ln ({\it l}/a)} = F_i\{p_k \} \,.
\eqno(9)
$$
These equations can be linearized near the fixed point
$$
p_i({\it l})=p^*_i \qquad \mbox{ (\,for\quad all\quad  {\it l\,})}
\eqno(10)
$$
and investigated by the standard methods of
linear algebra.  The ordinary phase transitions are described by
the saddle points of such equations. The simplest saddle point in
two parameter space (Fig.2) has the straight-line trajectories in
\begin{figure}
\centerline{\includegraphics[width=5.1 in]{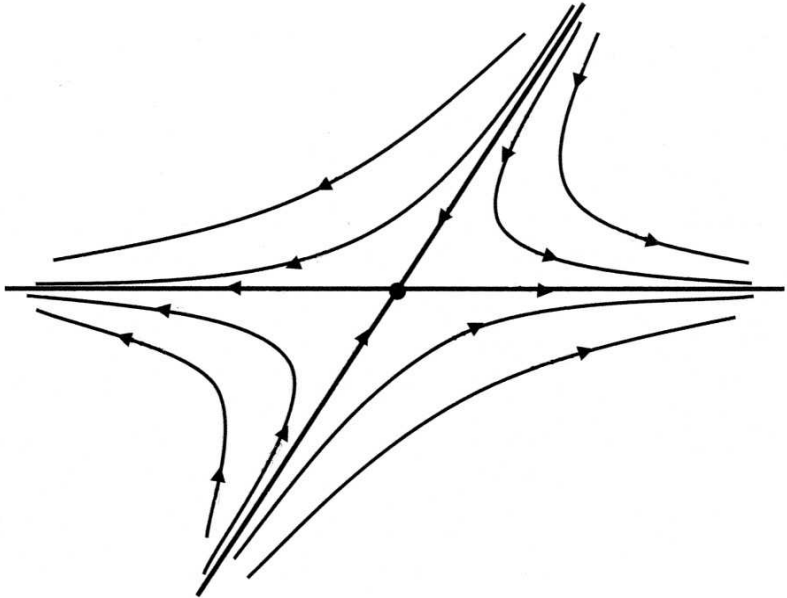}}
\caption{Simplest variant of the saddle point.} \label{fig2}
\end{figure}
two main directions (one stable and one unstable), while the rest
of trajectories are hyperbolic. For the usual phase transitions,
there are infinite number of stable directions and
one (in the simplest case)  unstable direction. The latter is
related with some controlling parameter like temperature,
measuring the distance to the critical point.

Instead of increasing  ${\it l}$ for fixed $a$, we can
diminish $a$ for fixed ${\it l}$.
The continuum limit  $a\to 0$
of field theory corresponds to
the critical surface $\xi/a=\infty$  in the
many-parameter space (Fig.3).
\begin{figure}
\centerline{\includegraphics[width=5.1 in]{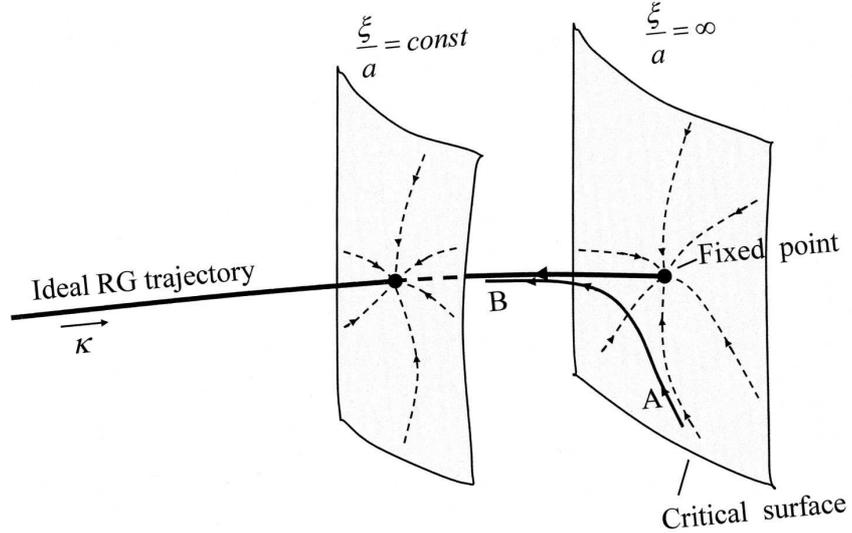}}
\caption{Schematic of the Wilson many-parameter space.}
\label{fig3}
\end{figure}
All trajectories at the critical
surface tend to the fixed point. The unstable trajectory,
originating in the fixed point will be referred as
an "ideal RG trajectory": along it one has the exact one-parameter
scaling, which is a pipe dream in many fields of physics
(see e.g. \cite{13a}).
 To define it
rigorously,
let us consider the limit $a\to 0$ with fixed
$\xi/a$;  then all trajectories lying at the surface
$\xi/a=const$ (Fig.3) tend to one point (analogously
to the critical surface), while
the locus of such points is the ideal RG trajectory.

Let the parameter $\kappa$ is  measuring the
distance along the
ideal trajectory: then $\xi/a$
(or $\Lambda/m$) is a function
of $\kappa$. Analogously, all dimensionless quantities
depend only on $\kappa$, while the dimensional quantities
are measured in units of $\Lambda$. As a result, we come
to equations
$$
g=F_g\left( \kappa\right)\,, \qquad
m=\Lambda F_m\left( \kappa\right) \,, \qquad
A_i=\Lambda^{d_i} F_i\left( \kappa\right)\,,
\eqno(11)
$$
which coincide with (5) and give the relations (6)
with no dependence on $m/\Lambda$.

The above construction has a following sense. If the limit
$a\to 0$ is taken in the arbitrary manner, then the
system will go to infinity along the unstable
direction and appear far from the critical surface,
which is our goal. Therefore, we suggest to take the continual
limit in two  steps:

(a) take a limit  $a\to 0$ for $a/\xi=const$;

(b) take a limit  $a/\xi\to 0$.

\noindent
It appears, that the dependence on $a$ in renormalized theory
disappears already at the first step.
The second step becomes
unnecessary and we need not take the continuum limit in the
bare theory. The $const$ appearing in (a) is one of the possible
definitions of the parameter $\kappa$.

These ideas are close to the QCD specialists, and in fact
the above consideration was
partially
taken from "Introduction to lattice QCD" by R.\,Gupta \cite{14}.
This picture is discussed there in relation to
improvement of the lattice action, and the author claims that
simulations, done along the ideal RG trajectory, will reproduce
the continuum physics without discretization errors. It implies
the absence of $a/\xi$ dependence, in accordance with our
results.  Only final conclusion was not made, that the continuum
limit is not necessary in the bare theory. In fact, this
conclusion goes across the present-day practice in lattice
simulations, which are made in the region of large $\xi/a$
(typically $\xi/a=5\div 15$) with accurate extrapolation to
$a/\xi\to 0$.

Any RG trajectory is a line of "constant physics", since
the RG transformation  is simply
the mental construction,
which does not affect the large-scale
properties of the system.
All trajectories belonging to the
critical surface and meeting in the fixed point  are
physically equivalent,  corresponding to the unique continuous
field theory.  The ideal RG trajectory originating in the fixed
point gives the equivalent representation for field theory.
 Let us consider the trajectory $AB$, which begins
near the critical surface and goes along it, and then
tends to the ideal RG trajectory (Fig.3).
Introducing  $\tilde\kappa$ as a distance along $AB$,
we come to the parametric representation analogous (11)
$$
g=\tilde F_g\left( \tilde\kappa\right)\,,
\qquad m=\Lambda \tilde F_m\left(\tilde \kappa\right) \,,
\qquad A_i=\Lambda^{d_i} \tilde F_i\left(\tilde\kappa\right)\,,
\eqno(12)
$$
and relations (6), independent of $\xi/a$. The choice of
small or large $\xi/a$ values  corresponds
to the "ends" of trajectory $AB$ which are arbitrary close
to the critical surface and the ideal trajectory; hence,
the obtained relations (6) correspond to continual theory.
However, the parametric
representation (12) is essentially different from (11) and is not
reduced to the change of variables $\kappa=f(\tilde\kappa)$.
To understand it,
let us retain definition of  $\kappa$ as a distance along the
ideal trajectory, and assign it to the point of $AB$,
corresponding to the same value of $\xi/a$. Then the second
relation (12) will be the same as (11), but the rest two
relations remain different:
$$
g=\tilde F_g\left( \kappa\right)\,,\qquad
m=\Lambda F_m\left( \kappa\right) \,,\qquad
A_i=\Lambda^{d_i} \tilde F_i\left(\kappa\right)\,.
\eqno(12')
$$
Indeed, the charge  $g$ usually belongs  to irrelevant
parameters and we can introduce "the axis of charges" at
the critical surface; the fixed point corresponds to $g=0$.
If the limit  $a/\xi\to 0$ is taken along the ideal trajectory,
then $g\to 0$. If this limit is taken along  $AB$, then
the arbitrary function $g=\tilde F_g(\kappa)$ in
$(12')$ is possible:  it depends on  the direction of  $AB$
relative to "the axis of charges". The functional relation
between $g$ and $\kappa$ becomes indeterminate and can be omitted.

As a result, the renormalized and bare sectors of theory become
decoupled. The renormalized sector contains relations
(6), where  $g$ and  $m$ are considered as independent variables.
The bare sector contains only relation
$a/\xi=m/\Lambda=F_m\left(\kappa\right)$, which determines
$\kappa$ as a function of $a$ and is irrelevant  from
viewpoint of physics. Parameter  $a/\xi$ becomes
absolutely free.

The set of different trajectories  $AB$ defines the universality
class of the corresponding field theory. Such trajectories  fill
the whole space, if the critical surface and the ideal RG
trajectory are unbounded.  In fact, the critical surface is
certainly restricted in some directions, because there are a lot
of such surfaces, corresponding to different phase transitions. To
obtain the correct relations  (6), there is no need to construct
the ideal RG trajectory: it suffices to find the arbitrary
trajectory like $AB$, belonging to the same universality class.

\vspace{2mm}

As a result, we come to the following conclusion:
\vspace{2mm}

\noindent
{\it Renormalizable theory of the considered type allows
representation in the form of  lattice theory, which gives
the correct
relations between physical quantities, and contains free
parameter $a/\xi$, which does not enter these relations. }

\vspace{3mm}
\begin{center}
{\bf 3. Confinement }
\end{center}

QCD with one sort of quarks contains two parameters, interaction
constant $g$ and the quark mass $m$. Its renormalization
properties are analogous to those of $\phi^4$ theory or QED and
are expressed by the relations (3,\,4); in fact Sec.2 set
axiomatic for study of such theories. We restrict our  discussion
by a theory without quarks, i.e. pure Yang-Mills theory; then  the
quark mass is not included  as a parameter and the theory does not
contain any natural mass scale. To avoid the specific difficulties
related with such situation, let us introduce the "extended
version" of Yang-Mills theory, where the role of the bare quark
mass $m_0$ (more exactly, the ratio $m_0/\Lambda$) is played by
some auxiliary parameter $p$ characterizing the lattice theory; as
a renormalized mass, we accept the mass  $m$ of the lightest
glueball (the bound state of several gluons), while the
correlation length $\xi$ is defined as $m^{-1}$. Thereby,  two
bare parameters  $g_0$ and $p$ provide the observable values for
renormalized $g$ and $m$ ($g$ corresponds to the momentum scale
$m$). In order to return to the standard variant of theory, we
should remove the introduced extra degree of freedom by fixing one
relation between observable quantities.  However, it can be done
on the late stage (see the end of Sec.3), while the main of
analysis is produced for the "extended version", which is
analogous to $\phi^4$ theory.

According to Wilson \cite{15}, confinement can be proved in the
lattice version of the Yang-Mills theory for large value of the
bare charge $g_0$. The energy of interaction for two probe
quarks, separated by a distance $R$, is $V(R)=\sigma R$, while the
string tension $\sigma$ and the glueball mass $m$ are given by
expressions  \cite{14,15,18}
$$
\sigma= \frac{\ln (3 g_0^2)}{a^2}
\,, \qquad m = \frac{4\,\ln (3 g_0^2)}{ a}
\,.
\eqno(13)
$$
In spite of the evident success, the Wilson theory is
considered as purely illustrative and having no relation
to real QCD. As was indicated by Wilson himself, his
theory corresponds to a situation
$$
\xi\ll a  \qquad \mbox{ or}\qquad  m\gg \Lambda \,,
\eqno(14)
$$
which is considered as unphysical. An attempt to advance into the
physical region inevitably destroys the strong coupling regime.
Indeed, fixing $\sigma$ to its observable value, we have
$g^2_0(a)=(1/3)\exp\{ \sigma a^2\}$ and substitution to the
Gell-Mann -- Low equation in the cut-off scheme \cite{16}
$$
-\frac{d\, g_0^2}{d \,\ln a^2} =\beta(g_0^2)=
          \beta_0 g_0^4+ \beta_1 g_0^6+\ldots
\eqno(15)
$$
gives  $\beta(g^2_0)=-g_0^2\ln( 3 g_0^2)$ for large
$g_0$ \cite{17}.
Together with a negative sign of $\beta_0$ and $\beta_1$
it implies the  negative $\beta$-function for all $g_0$ (Fig.4);
\begin{figure}
\centerline{\includegraphics[width=4.5 in]{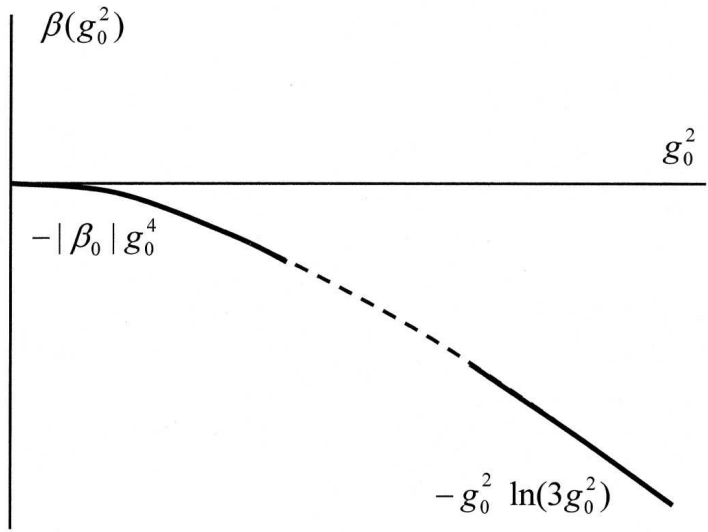}} \caption{The
Gell-Mann -- Low function of the Yang-Mills theory in the cut-off
scheme.} \label{fig4}
\end{figure}
the lattice results confirm this conclusion (see also \cite{29}).
In this case $g_0$ tends to zero in the continuum limit $a\to 0$.
It does not mean triviality of theory, because the behavior
$g_0\to 0$ is compatible with a finite value of the renormalized
charge $g$, as can be seen from the one-loop result for
$\Lambda\to\infty$
$$
g^2_0=\frac{g^2}{1+|\beta_0| g^2 \ln \Lambda^2/m^2} \,\,
\longrightarrow\,\, \frac{1}{|\beta_0|  \ln \Lambda^2/m^2}\,\,
\longrightarrow \,\,\, 0\,\,.
\eqno(16)
$$
Triviality is avoided, but the strong coupling
regime is inevitably destroyed and  Wilson's theory becomes
inapplicable.
\vspace{2mm}

The situation changes drastically, if we use representation
(5,\,6) introduced in the previous sections. In this case:

(1) We do not need to take the continual limit in the bare
theory, so $g_0$ remains finite.

(2) Due to absence of the $\xi/a$ dependence, this
parameter can be taken arbitrary: it eliminates objections
against the non-physical regime in Wilson's theory.

(3) Experience of $\phi^4$ theory shows that there is no direct
relation between the bare and renormalized
charge\,\footnote{\,Usually it is accepted that $g_0$
coincides with the renormalized charge $g$ taken at the scale
$\Lambda$; it is valid only if $g\ll 1$ and $g_0\ll 1$
simultaneously.}: representation
(5) is rigorously introduced in the limit $g_0\to\infty$
(see Footnote 2),
while $g$ remains to be a finite function of
$\kappa$ \cite{4}.
With some reservations, the same property is valid
in Yang-Mills theory.
Rewriting the second expression (13) in the form
$$
g_0^2=\frac{1}{3} \exp\left( \frac{m a}{4}\right) =
      \frac{1}{3} \exp\left\{ \frac{a}{4\xi} \right\}\,,
\eqno(17)
$$
we see that, independently of renormalized values of
$g$ and $m$, it is possible to  choose the free parameter $a/\xi$
so as to  obtain the sufficiently large value for $g_0$.  Then
Wilson's theory becomes applicable and the first relation (13)
gives finite value for  $\sigma$, i.e. confinement.

We have used the relations (13), which are valid for
the simplest Wilson action \cite{14,15,18}.
However, the latter is not suitable for our
purposes due to a trivial fact that it does not contain
the sufficient number of parameters.
To obtain the observable values
of $\sigma$ and $m$
$$
\sigma=a^{-2} f_\sigma(g_0)\,,\qquad
m=a^{-1} f_m(g_0)\,,\qquad
\eqno(21)
$$
one should fix both $g_0$ and $a$; but the fixed $a$ means that
it is impossible to introduce representation with free
parameter $a/\xi$. Therefore, we should consider some
generalizations.

The simplest Wilson action \cite{14,15,18}
$$
S=-\frac{1}{g_0^2} \sum_{\Box}  W_{\Box}^{1\times 1}
\eqno(19)
$$
is a sum over all plaquettes $\Box$
of size $1\times 1$, where the plaquette contribution
$W_{\Box}^{1\times 1}$ is determined by a product of
matrices attributed to the sides of a plaquette. In
the contemporary investigations, more complicated
forms of  the action are used
which contain contributions of $m\times n$ plaquettes
\cite{14}
$$
S=-\frac{1}{g_0^2} \sum_{\Box} \sum_{m,n} C_{mn}
W_{\Box}^{m\times n}   \,.
\eqno(20)
$$
The coefficients $C_{mn}$ sufficiently quickly decrease
with growth of $m$ and $n$.\,\footnote{\,To understand this
point, let us return to Eq.\,8. The exchange integrals $J_x$
should fall with $|x|$ in the exponential manner, in order
the bare spectrum $\epsilon(p)$ can be  regularly expanded
in $p$. Analogous arguments can be given for Yang-Mills
theory.  }
 If a contribution of the $n\times n$
plaquette is dominated in the sum, we obtain Eq.\,13
 with $na$ instead $a$. It is clear, that
generally we shall have the effective averaging of (13) over $a$
in some finite limits from  $a_{min}=a$ till $a_{max}=ka$.  As a
result, the relations (13) will have a form
$$
\sigma= \frac{\ln (3 g_0^2)}{a_1^2}
\,, \qquad m = \frac{4\,\ln (3 g_0^2)}{ a_2} \,,
\eqno(21)
$$
where  $a_1=k_1 a$, $a_2=k_2 a$ simply
by dimensional reasons.
These modifications do not
affect the qualitative conclusions made above.

The relation (6) for $\sigma$ has a form
$$
\sigma=m^2 F_\sigma(g)\,,
\eqno(22)
$$
so $g$ is functionally related with $\sigma/m^2$. Eqs.\,21 give
$$
\frac{\sigma}{m^2} = \frac{a_2^2}{ 16 a_1^2  \,\ln (3 g_0^2)} \,,
\eqno(23)
$$
and for $a_1\sim a_2$ the ratio $\sigma/m^2\,$ is small in
 the strong coupling region. It means that
only restricted range of $g$ values can be reproduced. Such
restriction is natural due to the physical essence of the problem.
Indeed,  the  linear confinement potential is expected  only at
large distances, where  $g$ is certainly not small; hence, small
values of  $g$ are inaccessible in the Wilson regime. On the
contrary, the restricted range of  $\sigma/m^2$ values goes across
the logic of theory. Indeed,  $a/\xi$ is a free parameter and all
physical results can be obtained (analytically or not)  at its
arbitrary value. In the case $a/\xi\gg 1$, the regime of
confinement is controlled analytically and any physically
accessible value of  $\sigma/m^2$ should be possible in this
limit. Probably, the range of $\sigma/m^2$ values can be extended
if we use the models with essentially different $a_1$ and
$a_2$.\,\footnote{\,Such models certainly exist. If contribution
of plaquette $1 \times n$ dominates in the sum of (20), then usual
tiling of the Wilson loop or correlational tube \cite{14,18} gives
$a_1^2=na^2$, $a_2=na$, and the right-hand side of Eq.23 is $n$
times greater than for the Wilson action. To understand which
values of  $\sigma/m^2$ are really accessible, it is necessary to
investigate, does the strong coupling regime still correspond to
condition $g_0\gg 1$ or it is replaced by the more general
$n$-dependent condition.}
%
Absence of restrictions on  $\sigma/m^2$ in the presence
 of restrictions on  $g$ is possible only
if $\sigma/m^2$  has a maximum as a function of $g$;
fortunately, we can demonstrate that it is really the case.

Investigations of more  complicated lattice versions of
Yang-Mills theory \cite{14} show existence of phase
transitions (lying in the region  $g_0\sim 1$), corresponding
to vanishing of the lightest glueball mass $m$, with finite
values  of  $\sigma$ and other mass parameters. These transitions
are considered as lattice artifacts, since they do not survive
in the continuum limit $a\to 0$, when  $g_0\to 0$.  In our
approach the limit  $a\to 0$ is not necessary and such phase
transitions acquire the physical sense. Their existence means
that the dependence  $\sigma/m^2= F_\sigma(g)$ is
singular  (Fig.5) and provides accessibility of arbitrary
$\sigma/m^2$ values,  retaining the restriction on
values of  $g$.
\begin{figure}
\centerline{\includegraphics[width=4.5 in]{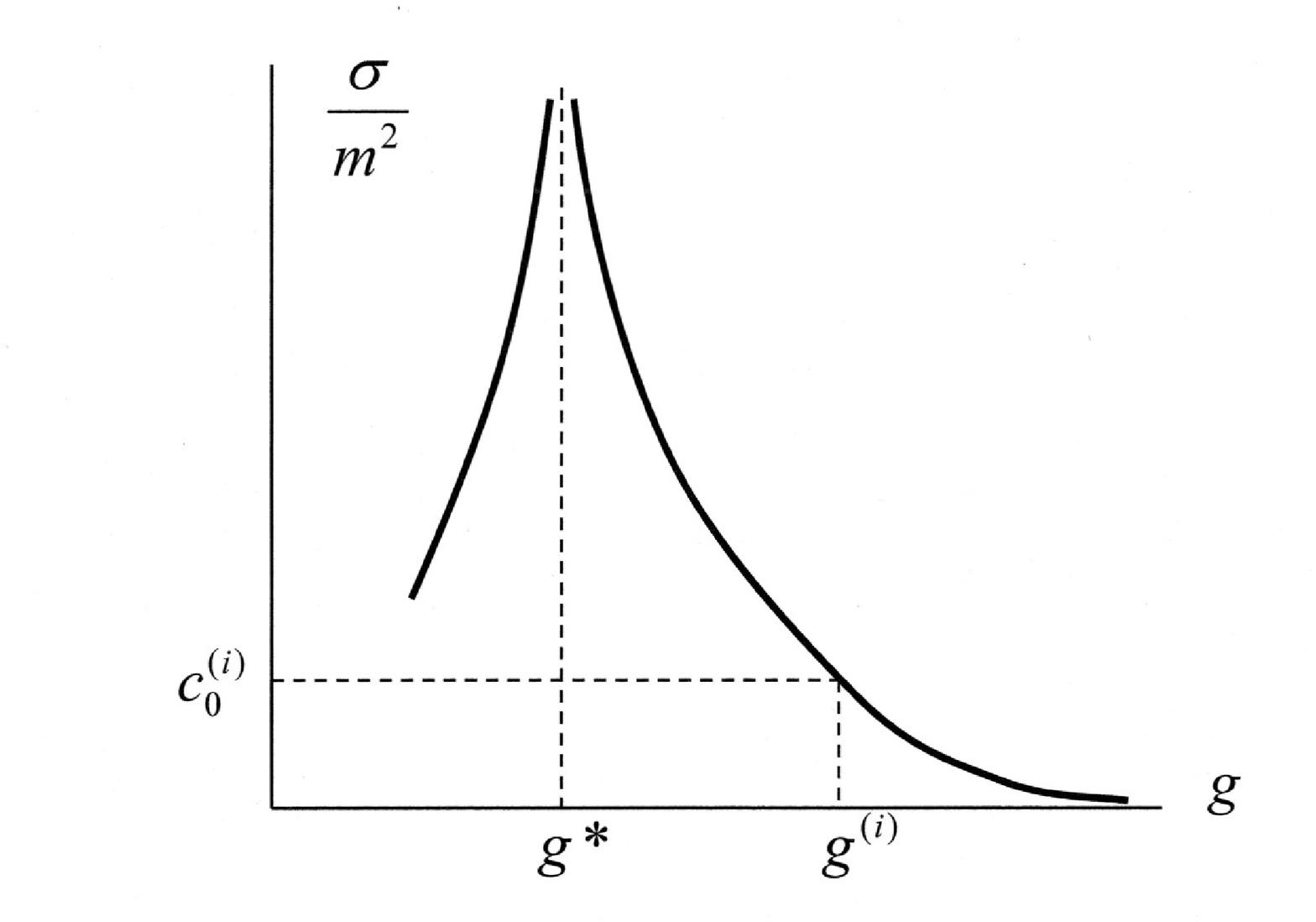}}
\caption{Dependence of $\sigma/m^2$ against $g$. In order to
obtain the special values $c^{(i)}_0$, corresponding to zero
values of the mass gap, one should mark all stable fixed points
$g^{(i)}$  on the horizontal axis and make a construction shown in
the figure.} \label{fig5}
\end{figure}

Existence of points
with $m=0$ in the parametric space means that
the "extended version" of Yang-Mills theory does not possess the
mass gap. To eliminate this defect, we should return to the
standard variant of theory, fixing one relation between
observable quantities. The character  of such relations
is well-known and is determined by the so called "dimensional
transmutation"  \cite[Sec.14.1]{14},
\cite[Sec.IV.6]{23}. If we have  $A=a^\mu f(g_0)$ for the
observable quantity $A$, then its independence of  $a$ means
$$
\frac{d A}{d a} =\mu a^{\mu-1} f(g_0)+
a^\mu \,\frac{d f(g_0)}{d g_0^2}\,\frac{d g_0^2}{d a}
= a^{\mu-1} \left[ \mu \,f(g_0) -2\, \frac{d f(g_0)}{d g_0^2}
\,\beta(g_0^2) \right]=0  \,,
$$
where Eq.15 is taken into account. Integration of the obtained
equation for  $f(g_0)$ gives
$$
A=const\, a^\mu \exp\left\{ \frac{\mu}{2}\, B(g_0^2)\right\}\,,
\qquad  B(g_0^2)=\int \,\frac{d g_0^2}{\beta(g_0^2)} \,,
$$
i.e. all quantities of the same dimensionality differ only
by the constant factor, independent of  $g_0$. For our purposes
it is convenient to accept the condition
$$
\sigma/m^2=c\,,
\eqno(24)
$$
which defines the one-parameter family of Yang-Mills
theories with different value of the structural constant
$c$.\,\footnote{\,"Extended" theory corresponds to the set
of all "standard" theories with different  $c$ values.}
Under condition  (24), the points with  $m=0$,
$\sigma=const$ become inaccessible.


It  does not yet prove the existence of a mass gap,
since $\sigma$ and $m$  can  vanish simultaneously.
In order to analyze such situations,
consider  the Gell-Mann -- Low equation for
the renormalized charge $g$ attributed to the scale  $m$
$$
\frac{d\, g^2}{d \,\ln m^2} =\beta(g^2)=
          \beta_0 g^4+ \beta_1 g^6+\ldots   \,,
\eqno(25)
$$
where  $\beta$-function does not coincide with  (15), but has
the same first coefficients  $\beta_0$ and $\beta_1$. It is clear
that value  $g^*$ (Fig.5) is a root of $\beta(g^2)$;
generally, it has several roots determining the RG fixed
points. In the limit $m\to 0$, the charge $g$ tends to one of
these fixed points, while following
variants are possible for $\sigma/m^2$:  (a) $\sigma/m^2 \to
\infty$, (b) $\sigma/m^2 \to 0$, (c) $\sigma/m^2 \to c_0$.
The first two variants are incompatible with Eq.24,
while the third variant is possible in the case  $c=c_0$. If
there are several stable fixed points  $g^{(i)}$, then there are
several special values  $c^{(i)}_0$ (see Fig.5), for which the
mass gap vanishes; for all other values of  $c$ the mass gap is
finite.

Physically, it looks most probable\,\footnote{\,Calculation
of $\beta$ functions in different theories \cite{5,29}
shows that they usually have the simple behavior interpolating
between strong coupling and weak coupling regime.}
 that only one fixed point
$g^*$ with $\sigma/m^2 \to \infty$ is present, so no special
values $c^{(i)}_0$ arise. Mathematically, one can suggest an
infinite number of fixed points, which form a sequence
$c^{(i)}_0$ everywhere dense in the interval $(0,\infty)$.
However, small values of  $\sigma/m^2$ correspond to
the Wilson regime where finiteness of  $\sigma$ and $m$ is
verified
immediately.
As a result, the proof of the
mass gap is complete for small values of the structural
constant  $c$.\,\footnote{\,In fact, we have
 suggested that the "extended" Yang-Mills
theory belongs to the type considered  in Sec.2. Motivation for
this is as follows. The bare Yang-Mills theory contains the single
parameter $g_0$, immediately related with the unstable direction.
We can extend theory along the stable directions in many-parameter
space; if there are unstable directions, we can artificially
forbid extension along them. Indeed, additional essential
parameters correspond to a theory, which is more general than
Yang-Mills theory; such theories certainly exist, but they are not
a subject for our consideration. We see that belonging of the
"extended" Yang-Mills theory to the type considered  in Sec.2 can
be accepted axiomatically. } The real perspective to strengthen
this statement is outlined in Footnote 6.

It is worthwhile to indicate the papers \cite{6,24},
which deal with  $\beta$-functions, close to (25).  The paper
 \cite{24} considers  $\beta(g^2)$ defined in the
 $MS$-scheme, where differentiation in  (25) is performed over
 arbitrary momentum scale  $\mu$;  behavior
$\beta(g^2)=\beta_\infty g^{2\alpha}$ with $\alpha\approx -13$ is
obtained for large  $g$, while the sign of $\beta_\infty$ remained
indefinite, so existence of fixed point is one of the possible
variants. Alternative definition of $\beta(g^2)$ can be obtained
in QCD, if $g$ is attributed to the scale of the quark mass $m$;
if $g$ is defined through the quark-gluon vertex, then calculation
of the asymptotics for $\beta$-function can be performed in a
complete analogue with QED \cite{6}, giving result $\beta(g^2)=
g^{2}$ with necessary existence of  a fixed point. We have seen
above the existence of fixed point when the glueball mass $m$ was
making the scale. The listed definitions of  $\beta(g^2)$ are
different technically, but physically correspond to the same
dependence of renormalized charge on the length
scale\,\footnote{\,According to \cite{25}, existence of the root
of the  $\beta$-function is invariant property, valid in all
physical renormalization schemes.}. The physical sense of
existence of fixed point was clarified above.

If massless quarks\,\footnote{\,In the case of fermions,
the mass renormalization is multiplicative and the choice
of the zero bare mass
provides zero value of the renormalized mass.}
are introduced, then the
regime of dimensional transmutation  is conserved and the
trick with "extension" of theory remains actual; it seems,
that the general structure of theory is also retained.

\vspace{2mm}

Our final conclusions  are as follows:
\vspace{2mm}

\noindent
{\it Whatever are properties of continuous  Yang-Mills theory,
there exists a lattice theory,  which reproduces them. The bare
charge $g_0$ in this lattice theory can be taken arbitrary,
and in particular infinitely large. Any reasonable
lattice version of Yang-Mills theory gives
finite values of $\sigma$ and $m$  in the strong
coupling limit. Vanishing of $\sigma$ and $m$  is possible
for exceptional configurations in many-parameter space, which
are avoided in the general situation.
 As a result, the problem of  analytical proof of
confinement and the mass gap
can be considered as solved, at least on the physical
level of rigor.  }

\end{document}